# Measuring the research output and performance of the university of Ibadan from 2014 to 2023:
## A scientometric analysis

**Dr. Muneer Ahmad[1]**
**&**
**Dr. Undie Felicia Nkatv[2]**
*[1]Chief University Librarian*
*The Iddi Basajjabalaba Memorial Library*
*Kampala International University*
*Box 20000, Ggaba Road, Kansanga, Uganda*
*muneerbangroo@gmail.com; ahmad.muneer@kiu.ac.ug*
*Mobile: +256709415657*

*[2]Ag. Deputy Librarian*
*The Iddi Basajjabalaba Memorial Library*
*Kampala International University, Box 20000*
*Ggaba Road, Kansanga, Uganda. Nkavfelicia*
*@gmail.com; felicia.undie@kiu.ac.ug*
*Mobile: +2348025668229*

**Abstract**
*This study employs scientometric methods to assess the research output and performance of the University of Ibadan from 2014 to 2023. By analyzing publication trends, citation patterns, and collaboration networks, the research aims to comprehensively evaluate the university's research productivity, impact, and disciplinary focus. This article's endeavors are characterized by innovation, interdisciplinary collaboration, and commitment to excellence, making the University of Ibadan a significant hub for cutting-edge research in Nigeria and beyond. The goal of the current study is to ascertain the influence of the university's research output and publication patterns between 2014 and 2023. The study focuses on the departments at the University of Ibadan that contribute the most, the best journals for publishing, the nations that collaborate, the impact of citations both locally and globally, well-known authors and their total production, and the research output broken down by year. According to the university's ten-year publication data, 7159 papers with an h-index of 75 were published between 2014 and 2023, garnering 218572 citations. Furthermore, the VOSviewer software mapping approach is used to illustrate the stenographical mapping of data through graphs. The findings of this study will contribute to understanding the university's research strengths, weaknesses, and potential areas for improvement. Additionally, the results will inform evidence-based decision-making for enhancing research strategies and policies at the University of Ibadan.*

## Introduction

The Nigeria's oldest degree-granting university is University of Ibadan (UI), founded in 1948 as an external College of the University of London and converted to an independent University in 1962. Located in Ibadan, the capital of Oyo State, UI has grown to become a premier institution in Nigeria and Africa, renowned for its academic excellence and research contributions **(Adebowale, 2015)**. The Arts, Science, Medicine, Agriculture, and Social Sciences are just a few of the faculties at UI that provide a broad range of undergraduate and graduate degrees. The university's strong academic standards and extensive curriculum, which are intended to educate students in a wide range of professional sectors, demonstrate its commitment to education (Adeniyi, 2019). UI has produced several eminent alumni over the years who have made major contributions to various areas worldwide. The university is home to the Kenneth Dike Library, one of the largest and most resourceful libraries in Nigeria. Named after the university's first African vice-chancellor, the library plays a critical role in supporting the institution's academic programs by providing access to vast collections of books, journals, and electronic resources **(Ogunruku, 2018)**. The library's extensive holdings and modern facilities make it an essential component of UI's academic environment.

Research is an important part of the University of Ibadan's mission. Among the research centres and institutions that are part of the school are the Institute of African Studies, the Center for Sustainable Development, and the Institute for Advanced Medical Research and Training. These centers support multidisciplinary research that tackles regional and global issues, making a substantial contribution to the corpus of knowledge across multiple disciplines **(Falola & Adebayo, 2016)**. The University of Ibadan's impact extends beyond academia. It plays a pivotal role in community development through various outreach programs and partnerships with local and international organizations. These initiatives aim to address social issues, improve public health, and enhance educational opportunities in the region **(Akinyemi, 2017)**. UI's engagement with the community underscores its commitment to societal development and its role as a catalyst for change. The university's campus spread over 1,000 hectares, provides a conducive environment for learning and research. It features modern classrooms, laboratories, and residential facilities, ensuring that students and staff have the necessary resources to excel. Additionally, the university's emphasis on extracurricular activities, including sports and cultural events, contributes to the holistic development of its students **(Olayinka, 2020)**. In conclusion, the University of Ibadan stands as a beacon of academic excellence in Nigeria and Africa. Its rich history, commitment to research, and active engagement with the community highlight its role as a leading institution dedicated to fostering knowledge and development. As UI continues to evolve, it remains focused on its mission to provide quality education and contribute to societal advancement.

**Literature Review**

In a research article, a review of the literature is crucial because it gives a thorough background that places the current topic within the framework of what is already known. It identifies gaps, trends, and key findings in the field, ensuring that the research is built on a solid foundation. By synthesizing relevant studies, the review helps to justify the research questions and objectives, demonstrating the study's significance and relevance. Moreover, it allows researchers to avoid redundancy, build upon previous work, and position their contributions within the broader academic discourse. In scientometric studies, the literature review is particularly crucial as it enables the identification of influential works, emerging topics, and collaboration patterns, ultimately guiding the direction and impact of the research.

Ahmad and Batcha (2020), titled "Lotka's Law and Authorship Distribution in Coronary Artery Disease Research in South Africa", examines the applicability of Lotka's Law to the authorship patterns in this field. Using publication data from the Web of Science for the period 1990-2019, the study analyzes 1,284 research papers authored by South African scientists. It applies bibliometric tools and statistical tests, including the Kolmogorov-Smirnov test, to assess the fit of Lotka's Law. The findings reveal that while the general distribution of authorship follows Lotka's inverse square law, the specific parameters for South African CAD research do not perfectly align, highlighting deviations in author productivity. The study underscores the importance of localized analysis in bibliometrics. Huang and Chang (2020) explored the research productivity of leading universities in the United States using scientometric indicators. Their study highlighted variations in publication output and citation impact across different institutions, with a focus on the influence of research funding and institutional support on academic performance. A bibliometric analysis of research performance in UK universities was carried out by Jones and Thomas (2019). According to the study, citation impacts and research visibility were higher at universities with robust international cooperation and interdisciplinary research projects. Gupta and Kumar (2021) used scientometric methods to evaluate the research productivity of Indian educational institutions. Their findings indicated significant growth in research output over the past decade, with notable increases in collaborative publications and international co-authorships. Lin and Chen (2018) performed a comparative scientometric study of research output in major Asian universities. They identified key research trends, institutional performance metrics, and collaboration patterns, revealing a growing emphasis on global research networks and high-impact publications. Gonzalez-Brambila et al. (2017) applied scientometric methods to evaluate the research performance of Latin American universities. Their study showed an increase in publication output and collaboration with international researchers, though with varying levels of citation impact across different institutions. Butler and Morrison (2016) conducted a bibliometric mapping study of research output in Australian universities. Their analysis revealed significant trends in research focus areas, collaboration networks, and the impact of institutional policies on research productivity. Al-Shehri (2020) performed a scientometric assessment of research productivity in universities across the Middle East. The study highlighted increasing research output and collaboration with international researchers, particularly in scientific and technical fields. Katz and Martin (2015) analyzed research collaboration patterns in universities using scientometric methods. Their study highlighted the positive impact of international and interdisciplinary collaborations on research productivity and impact. Ivanov and Popov (2016) performed a bibliometric review of research performance in Russian universities. The analysis indicated a growing trend in publication output but a need for improvement in citation impact and international collaboration. Demir and Aydin (2019) conducted a scientometric evaluation of research trends in Turkish universities. Their study revealed significant growth in research output and international collaborations, with variations in citation impact across different institutions. Lee and Choi (2021) compared research impact in universities from emerging and developed nations using scientometric methods. The study showed that universities in developed countries generally had higher citation impacts, though emerging nations were rapidly increasing their research output and international visibility. Andersson and Nilsson (2020) evaluated the research performance of universities in Scandinavia using scientometric indicators. The analysis revealed strong research productivity and high citation impacts, supported by robust institutional policies and funding. Bester and Van Staden (2016) performed a bibliometric study of research trends in South African universities. Their findings indicated an increase in research output and collaborations but highlighted ongoing challenges in achieving global research impact. Ahmad (2022) explored scientometrics investigation on coronary artery disease research among BRICS countries and has revealed that continuous rapid growth was found from 1990 to 2019. Also, the collaboration of country-wise and year-wise collaborative research is in an increasing trend. The majority of the work on coronary artery disease by scientists preferred to publish their research papers in journal articles. From the study, it has been revealed that the People's Republic of China contributed the highest productivity of articles and the lowest publications contributed by South Africa among the BRICS Countries for three decades. English is, by and large, the medium of research communication, for it is widely recognized worldwide. In the study, multiple authorship patterns are dominant compared to single author productivity in the BRICS Countries on coronary artery disease research publications. El-Khoury and Kassem (2018) reviewed research output and impact in higher education institutions across the Arab world using scientometric methods. The study revealed growth in research productivity and international collaborations, with varying levels of citation impact.

**Research Objectives**
- To look at the research output growth pattern of the University of Ibadan year sum 2014-2023.
- Discover which University of Ibadan authors are the most prolific.
- Determine which reputable journals publish academic works from the University of Ibadan.
- Ascertain which countries would be most suitable for collaborating on the publication of their study findings.
- To calculate publication density, the top 20 authors, cooperating nations, and institutions will be mapped based on the quantity of research articles they have authored.

**Research Methodology**
The current study measures the productivity of research at the University of Ibadan using the Clarivate Analytics Web of Science database. The data were retrieved from the Web of Science (WOS) database (https://www.webofscience.com/wos/woscc/basic-search) using a basic search. By searching for the term " University of Ibadan " in the affiliation field with the date span set to 2014–2023 and the indexes SCI–EXPANDED, SSCI, and AHCI, publications published by the University of Ibadan were found on WOS. A total of 7159 articles were located, and Histcite, VOS viewer, and MS-Excel software programs were used to examine the gathered data. Bibliometric tools and techniques have carefully examined the computed data to obtain the intended outcome that satisfies the study objectives.

**Results and Discussion**
**Table 1: Main information about the data**

| S. No. | Table I: Main Information about Data | No. of Publications 2014-2023 |
|---|---|---|
| 1 | Timespan | 2014:2023 |
| 2 | Sources (Journals, Books, etc.) | 2076 |
| 3 | Documents | 7159 |
| 4 | Annual Growth Rate % | -15.6 |
| 5 | Document Average Age | 4.66 |
| 6 | Average citations per doc | 30.53 |
| 7 | References | 243266 |
| | DOCUMENT CONTENTS | |
| 8 | Key words Plus (ID) | 11867 |
| 9 | Author's Keywords (DE) | 15138 |
| | AUTHORS | |
| 10 | Authors | 65646 |
| 11 | Authors of single-authored docs | 236 |

This assesses the bibliometric results and provides an overview of the most important bibliometric statistics. The published Web of Science (WoS) papers are grouped on Table I. A total of 7159 documents released between 2014 and 2023, make up the dataset. The sources of these materials include 2076 journals in addition to additional sources. With a yearly growth rate of -15.6 and an average age of 4.66 documents, the average number of citations each document received is also included in the dataset. It also gives the average annual number of citations per document, which is 30.53. Finally, the dataset includes the keywords used in the documents, including "Keywords plus (ID)" (11867) and "Author's Keywords (DE)" (15138). Utilizing this data, one can examine the trends and substance found in scientific papers. These statistics provide a comprehensive overview of the dataset's temporal scope, authorship dynamics, and document typology, offering a foundational context for the subsequent data analysis and findings.

**Table 2: Evaluation of the Annual Output of Publications of the University of Ibadan.**

| Table II: Annual Distribution of Publications and Citations | | | | | | | | |
|---|---|---|---|---|---|---|---|---|
| S. No. | Year | Records | % | Rank | TLCS* | % | TGCS** | % |
| 1 | 2014 | 434 | 6.06 | 10 | 457 | 9.13 | 8804 | 4.03 |
| 2 | 2015 | 448 | 6.26 | 9 | 664 | 13.27 | 11227 | 5.14 |
| 3 | 2016 | 524 | 7.32 | 8 | 718 | 14.35 | 29226 | 13.37 |
| 4 | 2017 | 598 | 8.35 | 7 | 735 | 14.69 | 32460 | 14.85 |
| 5 | 2018 | 649 | 9.07 | 6 | 766 | 15.31 | 41019 | 18.77 |
| 6 | 2019 | 664 | 9.28 | 5 | 518 | 10.35 | 24294 | 11.11 |
| 7 | 2020 | 857 | 11.97 | 4 | 491 | 9.81 | 38464 | 17.60 |
| 8 | 2021 | 986 | 13.77 | 2 | 384 | 7.68 | 18616 | 8.52 |
| 9 | 2022 | 951 | 13.28 | 3 | 202 | 4.04 | 10801 | 4.94 |
| 10 | 2023 | 1048 | 14.64 | 1 | 68 | 1.36 | 3661 | 1.67 |
| Total | | 7159 | 100.00 | | 5003 | 100.00 | 218572 | 100.00 |

*Total Local Citation Score, **Total Global Citation Score

In scientometrics, TGCS (Total Global Citation Score) and TLCS (Total Local Citation Score) are important citation-based metrics used to analyze the impact of research publications. It represents the total number of citations a particular paper has received from all sources indexed in a given database (e.g., Web of Science). TGCS score reflects the overall impact of a publication in the global research community. TLCS represents the number of times a paper has been cited by other papers within the same dataset or collection being analyzed.

Table II lists the total number of citations, average citations per year (TCpY), and number of papers produced by different authors each year. The data also contains the number of citations each author has gotten, the number of citations their publications have received, and the average number of citations each year. The preceding table shows the distribution of publications by year from 2014 to 2023. There was a divergence in the growth of publications starting in 2014. Ten years later, 1048 papers—out of 7159 total—have been published, the most noteworthy of which is 2023. The University of Ibadan published 7159 publications in a variety of areas between 2014 and 2023, according to Web of Science databases. Based on the study shown in Table II, there was a variance in the publication rate between 2014 and 2023. The year 2023 had the greatest rate, followed by 2021, 2022, and 2020. More than 60% of the articles were available throughout these four years, with 2023 making up the largest contribution with 7.55% of all publications. During this time, 7159 publications had 218572 citations found. There has been a discernible rise in the number of citations each year. The most citations occurred in 2023 (14.64%), 2021 (13.77%), and 2022 (13.28%) following closely behind. Citations had increased significantly over the preceding four years, as determined by calculating the citation growth rate.

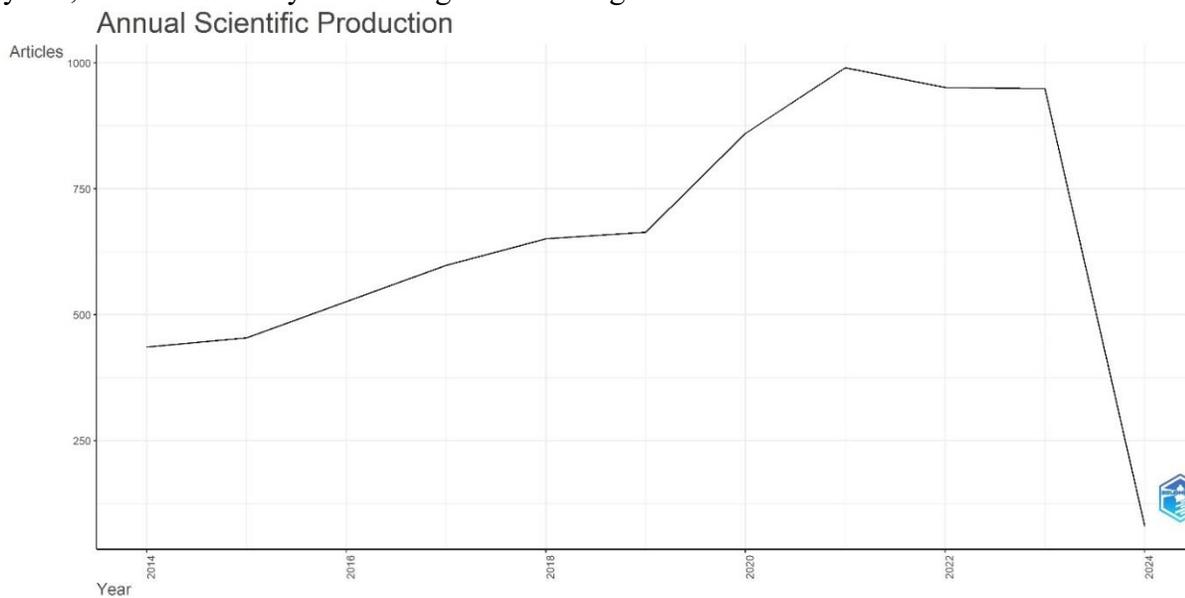

**Figure 1**

**Table 3: Analysis of the Publication Output of Top 20 Authors of the Universit of Ibadan**

| Table III: Publication Output of Top 20 Authors and Citation Score | | | | | |
|---|---|---|---|---|---|
| S.No. | h-index | Author | Citation sum within h-core | All citations | All articles |
| 1 | 75 | Mokdad AH | 78220 | 79378 | 108 |
| 2 | 73 | Fischer F | 74405 | 75427 | 101 |
| 3 | 70 | Yonemoto N | 79968 | 80982 | 97 |
| 4 | 68 | Shaikh MA | 72529 | 73501 | 94 |

| 5 | 68 | Jonas JB | 81066 | 81815 | 92 |
|---|----|----------|-------|-------|-----|
| 6 | 67 | Malekzadeh R | 84798 | 85132 | 78 |
| 7 | 67 | Sepanlou SG | 81826 | 82127 | 79 |
| 8 | 67 | Murray CJL | 75073 | 75780 | 89 |
| 9 | 64 | Dandona R | 70808 | 71351 | 81 |
| 10 | 64 | Mohammed S | 64005 | 64954 | 104 |
| 11 | 64 | Dandona L | 70808 | 71302 | 80 |
| 12 | 63 | Hay SI | 68203 | 68864 | 83 |
| 13 | 63 | Naghavi M | 68980 | 69513 | 79 |
| 14 | 62 | Vos T | 80695 | 81066 | 73 |
| 15 | 61 | Koyanagi A | 63919 | 64616 | 85 |
| 16 | 61 | Majeed A | 70207 | 70694 | 79 |
| 17 | 60 | Owolabi MO | 63266 | 64516 | 153 |
| 18 | 60 | Kim YJ | 64338 | 64910 | 77 |
| 19 | 60 | Rawaf S | 56087 | 56852 | 88 |
| 20 | 60 | Khader YS | 71025 | 71661 | 80 |

The table presents the publication output and citation scores of the top 20 authors, focusing on those affiliated with the University of Ibadan. Among them, Professor Mayowa O. Owolabi (h-index: 60, 153 articles, 63,266 citations within h-core) stands out as the only author from UI, demonstrating his significant scholarly impact. His high h-index and citation count reflect a strong research influence, particularly in the medical and health sciences.

Compared to other top authors, Owolabi has one of the highest publication counts (153 articles), indicating extensive research output. Although his total citations are slightly lower than some global counterparts, his consistent contributions make him a leading figure in UI's research landscape.

Overall, this ranking highlights globally influential researchers, with Owolabi representing UI among this elite group. His high citation impact and research productivity reinforce the university's strength in medical and public health research.

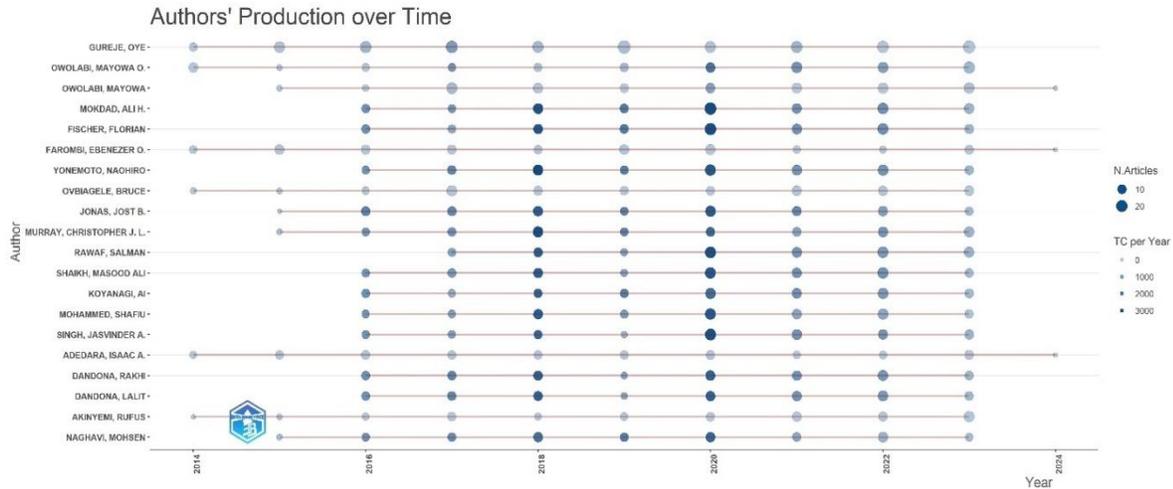

**Figure 2**

It looks at several different parameters to gauge the influence of different writers. A scientist or scholar's productivity and citation impact are gauged by their h-index. A scientist's productivity and impact are gauged by their g-index, which takes into consideration the quantity of highly referenced publications they have published. As indicators of an author's influence, the TC (total citations) and NP (number of publications) are also provided. The authors' lists were arranged in the table according to h-index, but it's vital to remember that different metrics might provide varying perspectives on an author's effect. As a general rule, it's best to take into account multiple metrics in order to obtain a full picture.

**Table 4: Analysis of Source-Wise Distribution of Documents**

| S.No. | Sources | Documents | h index | g index | m index | TC |
|---|---|---|---|---|---|---|
| | **Table IV: Top 20 Source –wise Document Ranking** | | | | | |
| 1 | Lancet | 86 | 66 | 86 | 6 | 74752 |
| 2 | PLOS One | 163 | 21 | 31 | 1.909 | 1621 |
| 3 | BMC Public Health | 71 | 18 | 30 | 1.636 | 1083 |
| 4 | Lancet Global Health | 42 | 18 | 42 | 1.636 | 3939 |
| 5 | African Health Sciences | 90 | 17 | 22 | 1.545 | 809 |
| 6 | American Journal of Tropical Medicine and Hygiene | 79 | 16 | 27 | 1.455 | 771 |
| 7 | Lancet Neurology | 19 | 16 | 19 | 2 | 17547 |
| 8 | BMC Pregnancy and Childbirth | 30 | 15 | 25 | 1.5 | 679 |
| 9 | BMJ Global Health | 30 | 15 | 26 | 1.875 | 704 |
| 10 | Environmental Science and Pollution Research | 51 | 15 | 24 | 1.5 | 669 |
| 11 | Psychological Medicine | 20 | 15 | 20 | 1.364 | 3479 |

| 12 | Reproductive Health | 26 | 15 | 26 | 1.364 | 948 |
|----|---------------------|-----|-----|-----|-------|-----|
| 13 | Scientific Reports | 55 | 15 | 23 | 1.875 | 686 |
| 14 | Resources Policy | 29 | 14 | 26 | 1.556 | 721 |
| 15 | Biomedicine & Pharmacotherapy | 19 | 13 | 19 | 1.625 | 607 |
| 16 | ANDROLOGIA | 26 | 12 | 20 | 1.091 | 447 |
| 17 | BMC Health Services Research | 37 | 12 | 21 | 1.2 | 490 |
| 18 | Clinical Infectious Diseases | 27 | 12 | 24 | 1.333 | 623 |
| 19 | Energy | 16 | 12 | 16 | 1.091 | 934 |
| 20 | Heliyon | 58 | 12 | 18 | 2 | 490 |

Table IV examines multiple metrics to quantify the influence of sources. A scientist or scholar's productivity and citation impact are gauged by their h-index. A scientist's impact and productivity are gauged by their g-index, which takes into consideration the quantity of highly referenced publications they have published. The TC (total citations) and NP (number of publications) are also included as metrics of an author's impact. The table provided lists the authors in descending order of h-index, with the Lancet having the highest h-index (66), followed by PLOS One with an h-index of (21) and Heliyon having the lowest (12). However, when looking at other metrics such as the g-index, the Lancet has the highest g-index (86) then followed by PLOS One, (31) and Energy has the lowest (16). It is important to note that different metrics can provide different perspectives on an author's impact, and it is usually recommended to consider multiple metrics to get a more complete picture.

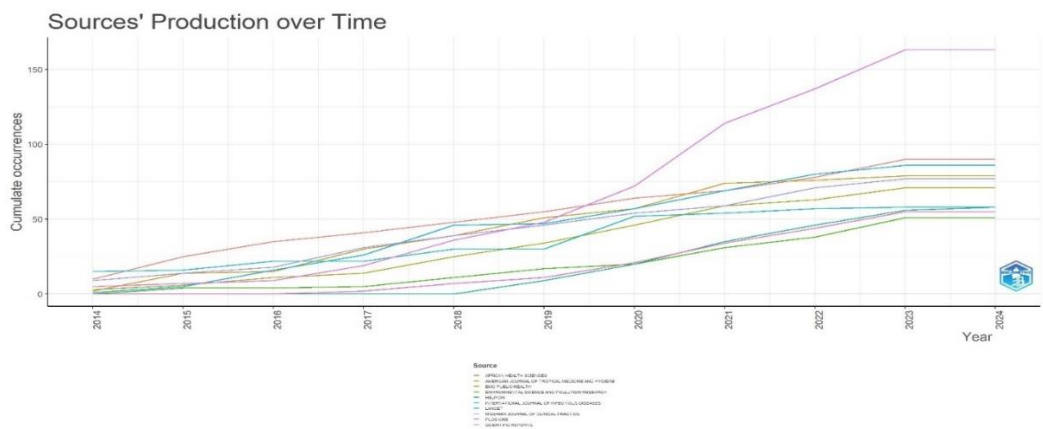

**Figure 3**
**Table 5: Analysis of Collaborative Ranking of Institutions**

| Table V: Ranking of Collaborative Institutions | | | | | |
|------|-----------------------------|---------|-------|-------|-------|--------|
| S.No. | Institution | Records | % | TLCS | TGCS | ACPP |
| 1 | University College Hospital | 1105 | 15.44 | 1011 | 80340 | 72.71 |
| 2 | Obafemi Awolowo University | 363 | 5.07 | 371 | 25078 | 69.09 |
| 3 | University Cape Town | 321 | 4.48 | 782 | 94791 | 295.30 |
| 4 | University Ilorin | 313 | 4.37 | 249 | 5340 | 17.06 |

| | | | | | |
|---|---|---|---|---|---|
| 5 | University Lagos | _310_ | 4.33 | 533 | 60668 | 195.70 |
| 6 | Ahmadu Bello University | _278_ | 3.88 | 815 | 71047 | 255.56 |
| 7 | London School of Hygiene & Tropical Medicine | _242_ | 3.38 | 627 | 80406 | 332.26 |
| 8 | University Witwatersrand | _232_ | 3.24 | 520 | 52993 | 228.42 |
| 9 | University Oxford | _224_ | 3.13 | 636 | 92839 | 414.46 |
| 10 | Federal Medical Centre | _223_ | 3.11 | 381 | 5068 | 22.73 |
| 11 | University Washington | _213_ | 2.98 | 645 | 102746 | 482.38 |
| 12 | UCL | _210_ | 2.93 | 624 | 95170 | 453.19 |
| 13 | University KwaZulu Natal | _204_ | 2.85 | 623 | 79126 | 387.87 |
| 14 | Kings College London | _198_ | 2.77 | 689 | 83444 | 421.43 |
| 15 | University Ghana | _198_ | 2.65 | 478 | 11228 | 56.71 |
| 16 | University Sao Paulo | _190_ | 2.58 | 640 | 95802 | 504.22 |
| 17 | Ministry of Health | _185_ | 2.56 | 572 | 78028 | 421.77 |
| 18 | Kwame Nkrumah University Science & Technology | _184_ | 2.54 | 544 | 30622 | 166.42 |
| 19 | University of Nigeria | _183_ | 2.6 | 136 | 5818 | 31.79 |
| 20 | University of Melbourne | _182_ | 2.6 | 653 | 93089 | 511.48 |

This table presents the top institutions collaborating with the University of Ibadan (UI) based on research publications. The University College Hospital (UCH) tops the list with 1,105 collaborative publications (15.44%), which is expected given its close ties to UI in medical and health research. Other Nigerian universities such as Obafemi Awolowo University (OAU), University of Ilorin, University of Lagos, and Ahmadu Bello University also feature prominently, indicating strong domestic academic partnerships. International collaborations are significant, especially with institutions like University of Cape Town, University of Oxford, University of Washington, and Kings College London.

The researchers identified academic collaboration in publications on disruptive innovation in research productivity of discipline. Table V shows the number of publications from the top twenty academic institutions. University College Hospital has the highest number of publications 1105 (15.44%), TLCS 1011 and TGCS 80340 and ACPP 72.71, followed by Obafemi Awolowo University with 363 records (5.07%), TLCS 371, TGCS of 25078 and ACPP of 69.09, thirdly University Cape Town with 321 records (4.48%), TLCS of 782 and TGCS of 94791 and an ACPP of 295.30, then the University Ilorin, has 313 records, (4.37%), TLCS of 249, TGCS of 5340 and ACPP of 17.06, and University Lagos has a record of 310 (4.33%), TLCS of 533. a TGCS of 60668 and an ACPP of 195.

**Table 6: Analysis of Ranking of Department-wise Distribution**

| Table VI: Ranking of Department-Wise Distribution | | | | | |
|---|---|---|---|---|---|
| S.No. | Institution with Sub-Division | Records | % | TLCS | TGCS | ACPP |
| 1 | University Ibadan, College Medicine | _1591_ | 22.2 | 1235 | 40508 | 25.46 |

| 2 | University College Hospital | 292 | 4.08 | 305 | 21191 | 72.57 |
|---|---|---|---|---|---|---|
| 3 | University Ibadan, Department of Chemistry | 268 | 3.74 | 282 | 4741 | 17.69 |
| 4 | University Ibadan, Faculty of Public Health | 199 | 2.78 | 110 | 8835 | 44.40 |
| 5 | University Ibadan, University College Hospital | 198 | 2.77 | 119 | 3418 | 17.26 |
| 6 | University Ibadan, Faculty of Pharmacy | 196 | 2.74 | 93 | 2135 | 10.89 |
| 7 | University College Hospital, Department of Medicine | 169 | 2.36 | 280 | 38525 | 227.96 |
| 8 | University Ibadan, Department of Medicine | 169 | 2.36 | 339 | 30941 | 183.08 |
| 9 | University Ibadan, Faculty of Veterinary Medicine | 169 | 2.36 | 119 | 1702 | 10.07 |
| 10 | University Ibadan, Department of Zoology | 163 | 2.28 | 171 | 2302 | 14.12 |
| 11 | University Ibadan, Department of Psychiatry | 154 | 2.15 | 172 | 3609 | 23.44 |
| 12 | University Ibadan, Department of Epidemiology & Medical Statistics | 144 | 2.01 | 124 | 7993 | 55.51 |
| 13 | University Ibadan, Department of Biochemistry | 122 | 1.7 | 141 | 1717 | 14.07 |
| 14 | University Ibadan, Department of Community Medicine | 117 | 1.63 | 173 | 21869 | 186.91 |
| 15 | University Ibadan, Institute for Advanced Medical Research & Training | 111 | 1.55 | 302 | 43435 | 391.31 |
| 16 | University Ibadan, Department of Physics | 107 | 1.49 | 56 | 1060 | 9.91 |
| 17 | University Ibadan, Department of Geology | 104 | 1.45 | 16 | 763 | 7.34 |
| 18 | University Ibadan, Faculty of Science | 102 | 1.42 | 35 | 1186 | 11.63 |
| 19 | University Ibadan, Department of Microbiology | 101 | 1.41 | 57 | 2128 | 21.07 |
| 20 | University Ibadan, Department of Health Promotion & Education | 91 | 1.27 | 165 | 18493 | 203.22 |

This table presents an analysis of research output by department at the University of Ibadan (UI). The University of Ibadan's research landscape is strongly dominated by the medical and health sciences, as evidenced by the significant contribution of the College of Medicine, which accounts for 1,591 publications (22.2%). The University College Hospital (UCH) also appears multiple times in the ranking, reinforcing UI's strong focus on clinical and medical research.

The most impactful research departments based on average citations per paper (ACPP) include the Institute for Advanced Medical Research & Training (ACPP: 391.31), Department of Medicine (ACPP: 183.08), and the UCH Department of Medicine (ACPP: 227.96). While some departments have high research output, their citation impact remains relatively low. For example, the Department of Chemistry (268 papers, ACPP: 17.69) is among the top contributors in volume but has a lower citation impact, possibly due to a focus on local research topics with limited global visibility.

**Table 7: Distribution of Papers by Types of Documents**

| Table VII: Document Type Contribution of Research | | | | | |
|---|---|---|---|---|---|
| S. No. | Document Type | Records | % | TLCS | TGCS |
| 1 | Article | 5452 | 76.16 | 4519 | 191573 |
| 2 | Meeting Abstract | 846 | 11.82 | 6 | 146 |
| 3 | Review | 460 | 6.43 | 340 | 22974 |
| 4 | Editorial Material | 183 | 2.56 | 81 | 3052 |
| 5 | Letter | 63 | 0.88 | 33 | 254 |
| 6 | Article; Early Access | 43 | 0.60 | 0 | 66 |
| 7 | Correction | 35 | 0.49 | 0 | 18 |
| 8 | Book Review | 27 | 0.38 | 0 | 1 |
| 9 | Article; Proceedings Paper | 20 | 0.28 | 22 | 263 |
| 10 | Poetry | 5 | 0.07 | 0 | 2 |
| 11 | Biographical-Item | 4 | 0.06 | 0 | 0 |
| 12 | News Item | 4 | 0.06 | 1 | 32 |
| 13 | Review; Early Access | 4 | 0.06 | 0 | 12 |
| 19 | Letter; Early Access | 3 | 0.04 | 0 | 0 |
| 14 | Editorial Material; Early Access | 3 | 0.04 | 0 | 1 |
| 15 | Article; Book Chapter | 1 | 0.01 | 0 | 34 |
| 16 | Article; Data Paper | 1 | 0.01 | 0 | 2 |
| 17 | Book Review; Early Access | 1 | 0.01 | 0 | 0 |
| 18 | Fiction, Creative Prose | 1 | 0.01 | 0 | 0 |
| 20 | Review; Book Chapter | 1 | 0.01 | 1 | 141 |
| 21 | Meeting | 1 | 0.01 | 0 | 0 |
| 22 | Review; Retracted Publication | 1 | 0.01 | 0 | 1 |
| | Total | 7159 | 100 | 5003 | 218572 |

Table VII displays the type of document authors preferred in their research at the University of Ibadan. Out of the total 218572 documents, the maximum, i.e. 5452 (76.16%) was the journal articles and more preferred document type, followed by meeting abstracts with 846 (11.82%), review with 460 (6.443%), editorial material with 183 (2.56%), letters with 63 (0.88%), article early access with 43 (0.60%), correction with 35 (0.49%), Book review with 27 (0.38%), with the article, proceedings paper 20 (0.28) while the minor document type records at (0.01%), which comprise of Item like Article; Book Chapter; Book Review; Early Access Retracted Publication and Retraction; this record reveals Article has the highest document type contribution of research. This shows that 76.16% of the researchers frequently like article-type documents.

**Table 8: Most Relevant Countries by Corresponding Authors**

| | Table VIII: Most Relevant Countries by Corresponding Authors | | | | |
|---|---|---|---|---|---|
| S.No. | Country | Articles | Articles % | SCP | MCP | MCP % |
| 1 | Nigeria | 3753 | 52.4 | 2340 | 1413 | 37.6 |
| 2 | USA | 646 | 9 | 1 | 645 | 99.8 |
| 3 | United Kingdom | 371 | 5.2 | 0 | 371 | 100 |
| 4 | South Africa | 241 | 3.4 | 2 | 239 | 99.2 |
| 5 | China | 129 | 1.8 | 0 | 129 | 100 |
| 6 | Germany | 104 | 1.5 | 0 | 104 | 100 |
| 7 | Canada | 99 | 1.4 | 1 | 98 | 99 |
| 8 | Australia | 90 | 1.3 | 0 | 90 | 100 |
| 9 | Ghana | 54 | 0.8 | 0 | 54 | 100 |
| 10 | India | 54 | 0.8 | 0 | 54 | 100 |
| 11 | Ethiopia | 53 | 0.7 | 0 | 53 | 100 |
| 12 | Netherlands | 51 | 0.7 | 0 | 51 | 100 |
| 13 | Italy | 49 | 0.7 | 0 | 49 | 100 |
| 14 | Switzerland | 48 | 0.7 | 0 | 48 | 100 |
| 15 | Brazil | 45 | 0.6 | 1 | 44 | 97.8 |
| 16 | Benin | 40 | 0.6 | 15 | 25 | 62.5 |
| 17 | Kenya | 35 | 0.5 | 0 | 35 | 100 |
| 18 | Cameroon | 28 | 0.4 | 0 | 28 | 100 |
| 19 | Norway | 23 | 0.3 | 0 | 23 | 100 |
| 20 | Spain | 23 | 0.3 | 0 | 23 | 100 |

The countries with the highest number of publications and associated authors are listed in Table VIII. The SCP and MCP metrics, respectively, count the number of single corresponding authors and multiple corresponding authors. The MCP ratio and frequency shed information on the contributions to the field made by writers from different countries. Additionally, it displays the distribution of the most pertinent nations according to the relevant writers. Twenty countries' worth of output was examined in terms of articles, articles percentages, SCP, MCP, and MCP percentages. With 3753 papers (52.4%) published between 2014 and 2023, Nigeria has the most publications followed by the USA with 646 articles (9%), the UK with 371 pieces (5.2%), and South Africa with 241 articles (3.4%).

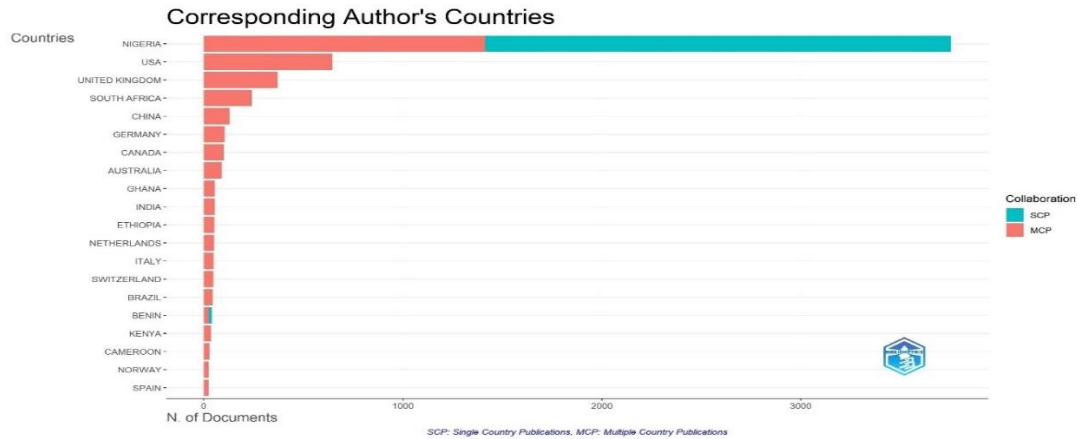

**Figure 4**

As Figure 4 illustrates, Nigeria and the United States as have the highest number of corresponding authors. The United States has 646 papers with corresponding authors, while Nigeria has 3753 articles. This indicates that research in both countries has MCP ratios that are greater than normal, indicating a higher likelihood of multiple related authorships for authors from both countries. The list emphasizes how multidisciplinary research is. The countries on the list are from many continents; for instance, South Africa, Canada, China, Germany, India, Australia, Brazil, and Ghana all have writers that match. This demonstrates the broad spectrum of research, from medical applications. Therefore, this list of nations and their corresponding authors sheds light on the worldwide significance of collaborative research. It also demonstrates the multidisciplinary nature of research and the need for international collaboration among scholars to investigate its possible uses.

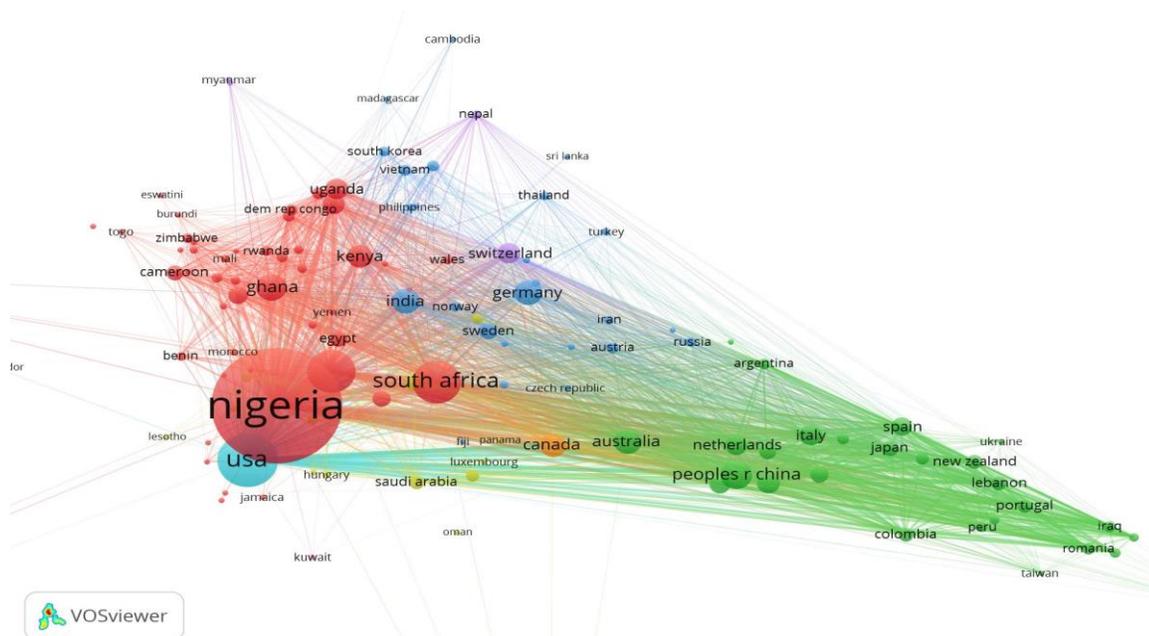

**Figure 5**

As we can see in Figure 5, the network of countries consists of linkages and clusters. The map's various hues represent the variety of study directions. The connections between nodes show a cooperative relationship between groups, whereas the larger nodes reflect the dominating nation. The degree of cooperative relationships among the countries is shown by the distance between the nodes and the thickness of the links. The three countries that contributed most significantly to research at the University of Ibadan were Nigeria, the US, and the UK. The fourth most influential nation in the creation of a cooperative network was South Africa. To promote future international cooperation and information exchange, it appears that the policies fostering cooperation with these nationally key countries need to be changed.

## Conclusion

This article covers the current study which is exclusive to the University of Ibadan. Numerous authors from different countries collaborated with the University of Ibadan to publish their articles during the study period. The data shows the number of articles published by each author. The data shows that the most relevant author is Mokdad AH, with 78220 Citations sum with h-core, 108 articles published, and an h-index of 75, this is followed by Fischer F, with 74405 citation sum with h- score, 101articles published, and an h-index of 73, and Yonemoto N with 79968 Citations sum with h-core, with an h-index 70 and 97 articles published. The other authors on the list include Shaikda MA, Jonas JB, and Malekzadeh R. This data indicates that these authors are leading researchers and have published many articles. The highest number of publications that could be found in 2023 was 1048, (14.64%). the second-highest number with 986 publications in 2011, which makes up (13.77%) of the University of Ibadan, and it was followed by the year 2022 with 951 articles (13.28%) during the study period. According to the study, the College of Medicine has contributed the maximum, i.e., 1591 (22.2%) publications.